\documentclass[conference,a4paper]{APSIPA2025}
\usepackage{amsmath}
\usepackage{graphicx}
\usepackage{multirow}
\usepackage{threeparttable}
\usepackage[backend=biber,style=ieee,]{biblatex}
\usepackage{amsmath}  
\usepackage{amssymb}  
\usepackage{bm}  
\usepackage{pifont} 
\usepackage{adjustbox}  
\usepackage{graphicx}   
\usepackage{geometry}
\usepackage{subcaption} 
\usepackage{fancyhdr}

\fancypagestyle{firststyle}{
  \fancyhf{}
  \fancyhead[C]{2025 Asia Pacific Signal and Information Processing Association Annual Summit and Conference (APSIPA ASC)}
}

\begin{document}

\title{A Distilled Low-Latency Neural Vocoder with Explicit Amplitude and Phase Prediction }

\author{
\authorblockN{
Hui-Peng Du, Yang Ai$^*$, and Zhen-Hua Ling
}

\authorblockA{
National Engineering Research Center of Speech and Language Information
Processing, \\University of Science and Technology of China, Hefei, P. R. China\\
E-mail: redmist@mail.ustc.edu.cn, yangai@ustc.edu.cn, zhling@ustc.edu.cn}}

\maketitle
\thispagestyle{firststyle}
\pagestyle{fancy}

\begin{abstract}
The majority of mainstream neural vocoders primarily focus on speech quality and generation speed, while overlooking latency, which is a critical factor in real-time applications.
Excessive latency leads to noticeable delays in user interaction, severely degrading the user experience and rendering such systems impractical for real-time use.
Therefore, this paper proposes DLL-APNet, a Distilled  Low-Latency neural vocoder which first predicts the Amplitude and Phase spectra explicitly from input mel spectrogram and then reconstructs the speech waveform via inverse short-time Fourier transform (iSTFT). 
The DLL-APNet vocoder leverages causal convolutions to constrain the utilization of information to current and historical contexts, effectively minimizing latency.
To mitigate speech quality degradation caused by causal constraints, a knowledge distillation strategy is proposed, where a pre-trained non-causal teacher vocoder guides intermediate feature generation of the causal student DLL-APNet vocoder. 
Experimental results demonstrate that the proposed DLL-APNet vocoder produces higher-quality speech than other causal vocoders, while requiring fewer computational resources.
Furthermore, the proposed DLL-APNet vocoder achieves speech quality on par with mainstream non-causal neural vocoders, validating its ability to deliver both high perceptual quality and low latency.
\end{abstract}

\section{Introduction}

\renewcommand{\thefootnote}{}
\footnote{$^*$ Corresponding author. This work was funded by the Anhui Province Major Science and Technology Research Project under Grant S2023Z20004, the National Nature Science Foundation of China under Grant 62301521 and the Anhui Provincial Natural Science Foundation under Grant 2308085QF200.}
 \renewcommand{\thefootnote}{\arabic{footnote}}
\addtocounter{footnote}{-1}
Neural vocoders convert input acoustic features (e.g., mel spectrogram) into speech waveform by neural networks, thus they directly impact the quality of synthesized speech and have found applications in various domains such as text-to-speech (TTS) \cite{kong2020hifi,siuzdakvocos}, speech enhancement (SE) \cite{mira2023voce}, voice conversion (VC) \cite{cao2024neuralvc}, etc.  

Early neural vocoders, e.g., WaveNet \cite{oord2016wavenet} and SampleRNN \cite{mehri2016samplernn}, employed auto-regressive methods to generate time-domain speech waveforms, which precluded parallel processing. 
In subsequent work, two categories of neural vocoders emerged: flow-based vocoders \cite{prenger2019waveglow,ping2020waveflow} and diffusion-based vocoders \cite{nguyen2024fregrad,koizumi2022specgrad}. 
While they can produce high-quality speech, their practical deployment, especially on devices with limited computing resources, remains challenging. 
Recently, generative adversarial networks (GANs) \cite{goodfellow2014generative} have demonstrated remarkable performance in generative tasks. 
GAN-based vocoders \cite{kong2020hifi,siuzdakvocos,ai2023apnet} are among the most prevalent vocoder architectures due to their relatively straightforward algorithmic design, which implicitly incorporates waveform quality improvements through discriminator supervision.

However, beyond the traditional focus on speech quality and generation speed, latency remains a critical yet often overlooked metric in most vocoder research, especially for real-time practical applications.
Many existing vocoders rely on standard convolutions, which cause the receptive field to expand as network depth increases when stacking convolutional blocks \cite{ai2024low}. 
This results in considerable model-intrinsic latency, making it infeasible to support frame-by-frame, concurrent transmission and computation—an unacceptable limitation in real-time systems from the receiver's perspective.
For example, in communication scenarios involving speech codecs \cite{li2024single,langman2024spectral} that transmit mel spectrograms, causal vocoders are essential to minimize latency and ensure real-time synthesis at the receiving end.

To address this, we propose DLL-APNet, a Distilled Low-Latency neural vocoder with explicit Amplitude and Phase spectrum prediction. 
Building on our previous work, APNet2 \cite{du2023apnet2}, the proposed vocoder introduces specific improvements to significantly reduce latency. 
In speech coding fields \cite{wu2023audiodec,zeghidour2021soundstream}, causal convolutions are commonly employed to reduce model latency and meet communication requirements. 
Therefore, we replace standard convolutions in APNet2 with causal convolutions, which use asymmetric padding to ensure the convolution kernel does not access future information, thereby minimizing computational latency. 
To compensate for the speech quality degradation caused by the introduction of causal convolutions, we adopt knowledge distillation training strategy, using a pre-trained APNet2 as the teacher model to guide the learning of the student model DLL-APNet. 
This allows DLL-APNet to extract maximal knowledge within its limited receptive field, facilitating the generation of high-quality speech waveform. 
As the results of our experiment, the speech quality synthesized by our proposed DLL-APNet outperforms other causal vocoders and remains comparable to non-causal models, indicating that the knowledge distillation strategy can mitigate the speech quality degradation caused by causal convolutions to some extent.

The rest of the paper is structured as follows. We introduce related works about GAN-based neural vocoders and low-latency speech generation methods in Section \ref{sec:rw} and describe our proposed DLL-APNet in Section \ref{sec:pm}. 
The experimental setups and results are presented in Section \ref{sec:es} and \ref{sec:ra}, respectively. 
Finally, we give conclusion in Section \ref{sec:c}.

\begin{figure*}
    \centering
    \includegraphics[width=0.8\linewidth]{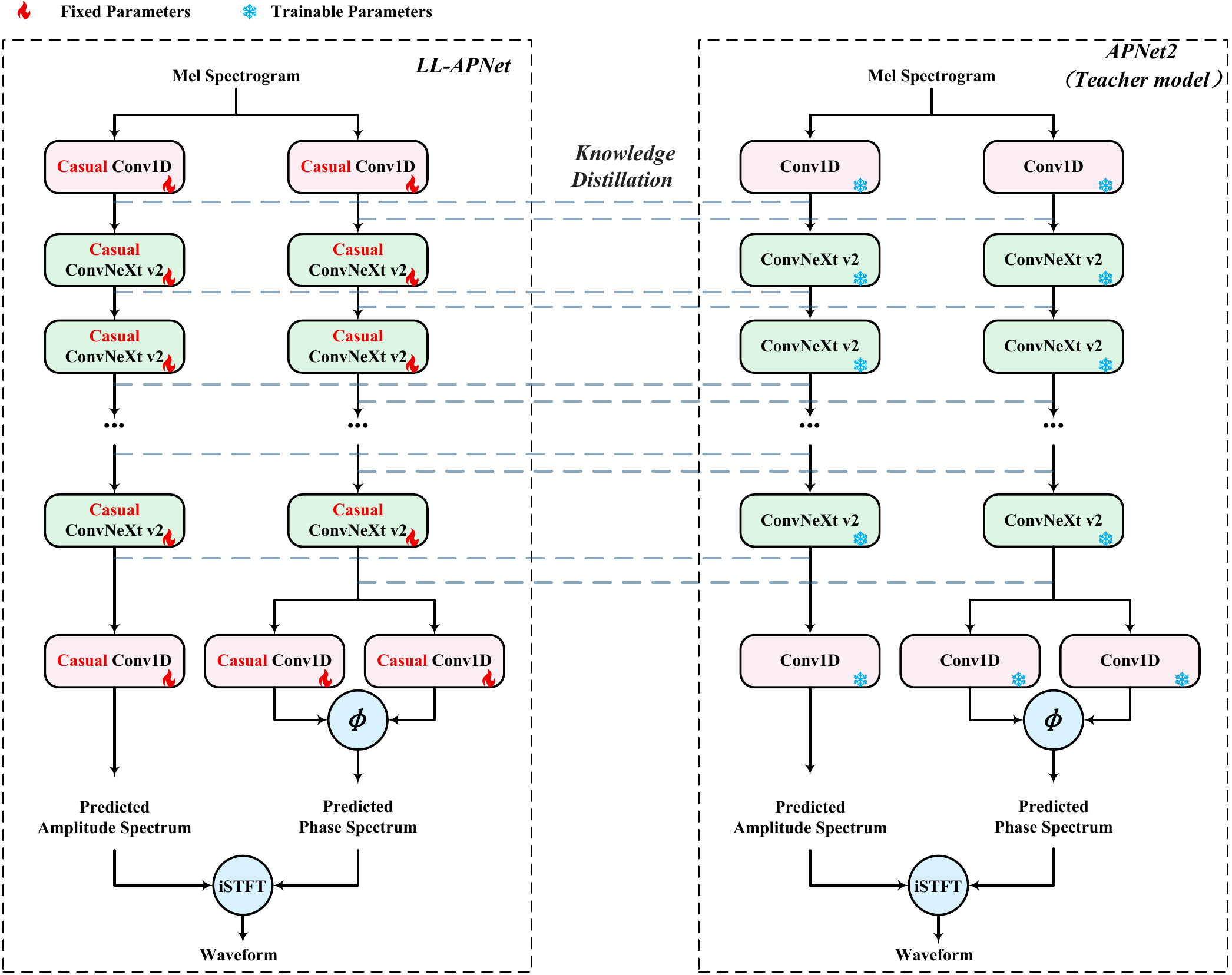}
    \caption{Model structure and knowledge distillation strategy of the proposed DLL-APNet vocoder, where $\phi$ denotes the phase calculation formula. During training, part of the parameters of DLL-APNet are trained through knowledge distillation from teacher model APNet2.}
    \label{fig:1}
\end{figure*}

\section{Related Work}
\label{sec:rw}
Since the benchmark employed in this study utilizes GAN-based neural vocoders and prioritizes causal modeling to reduce latency, this section offers a concise review of GAN-based neural vocoders and recent approaches to low-latency speech generation.
\subsection{GAN-based Neural Vocoders}
GAN-based neural vocoders represent one of the most dominant frameworks in modern vocoding technique, typically composed of a generator and a discriminator. 
We can classify GAN-based neural vocoders into non-all-frame-level and all-frame-level types based on whether upsampling operations are adopted. 
For non-all-frame-level neural vocoders (e.g., BigVGAN \cite{lee2022bigvgan}, HiFi-GAN \cite{kong2020hifi}, and iSTFTNet \cite{kaneko2022istftnet}), mel spectrograms are either left unprocessed or first upsampled to a higher temporal resolution before being directly converted into time-domain waveforms through transposed convolutional layers.
However, upsampling operations pose challenges for deployment on devices that lack parallel computing capabilities, as they introduce sequential dependencies and computational overhead. 
In contrast, all-frame-level neural vocoders (e.g., APNet \cite{ai2023apnet}, APNet2 \cite{du2023apnet2}, and Vocos \cite{siuzdakvocos}) predict amplitude and phase spectrum at the same frame rate as the input mel spectrogram through convolutional operations. 
The speech waveform is finally reconstructed via the inverse short-time Fourier transform (iSTFT), leveraging the spectral information to synthesize natural-sounding speech. 
Our previously proposed APNet vocoder \cite{ai2023apnet} achieved explicit  amplitude and phase spectrum prediction with small frame shifts for speech waveform reconstruction, especially  leveraging parallel estimation architecture and anti-wrapping loss function for accurate phase estimation. 
Our previously proposed APNet2 vocoder \cite{du2023apnet2} advanced APNet by integrating ConvNeXt v2 blocks \cite{woo2023ConvNeXt} and improved discriminators, enabling high-sampling-rate spectral prediction with large frame shifts. 
However, these approaches all overlooked latency issues, resulting in substantial model delays.

\subsection{Low-Latency Speech Generation Methods}
Latency, defined as the minimum amount of input time required to initiate the model, is a critical factor in many speech generation systems. 
It determines whether such models can be applied in real-time scenarios, such as speech communication.
Causal models exhibit extremely low model latency, with relevant research conducted in several low-latency speech generation methods, e.g., speech coding \cite{ai2024apcodec,wu2023audiodec,zeghidour2021soundstream}, VC \cite{ning2023dualvc}, SE \cite{tsunoo2025causal} and speech phase prediction \cite{ai2024low}. 
One approach \cite{ai2024apcodec} is to replace standard convolutions with linear layers and employ knowledge distillation strategies to enhance model performance. However, linear layers only allow the current output to access the current input rather than past inputs, presenting certain limitations. Another approach \cite{wu2023audiodec,zeghidour2021soundstream} is to use causal convolutions instead of standard convolutions and employ knowledge distillation strategies, avoiding the issue where standard convolutions with symmetric padding can access future information. 
However, low-latency neural vocoders have not yet been thoroughly investigated.

\section{Proposed Method}
\label{sec:pm}
An overview of the proposed DLL-APNet vocoder is shown in Figure \ref{fig:1}. 
Building upon our previously proposed APNet2 vocoder \cite{du2023apnet2}, the input mel spectrogram is processed by a amplitude spectrum prediction branch and a phase spectrum prediction branch in parallel to explicit predict the amplitude and phase spectra, respectively. 
The time-domain waveform is then reconstructed via iSTFT. 
To support low latency, all convolutional layers in the DLL-APNet vocoder are causal. 
At the training stage, a pre-trained APNet2 vocoder is used as a teacher model to generate intermediate features, which guide the distillation-based training of the proposed DLL-APNet vocoder. 
Details of the model architecture and training procedures will be presented below. 

\begin{figure}[t]
    \centering
    \begin{subfigure}[t]{0.95\linewidth}
        \includegraphics[width=\linewidth]{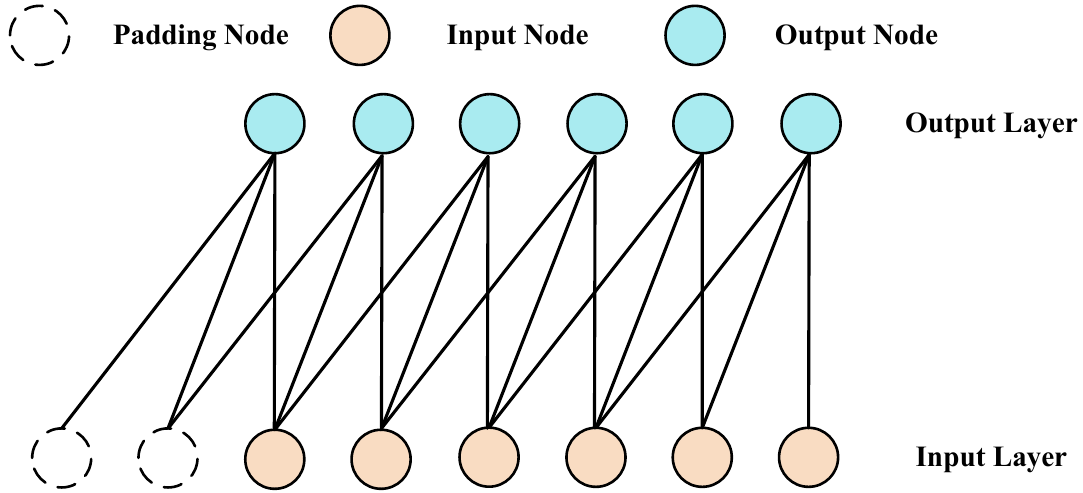}
        \caption{causal convolutional layer.}
    \end{subfigure}
    \hfill  
    \begin{subfigure}[t]{0.95\linewidth}
        \includegraphics[width=\linewidth]{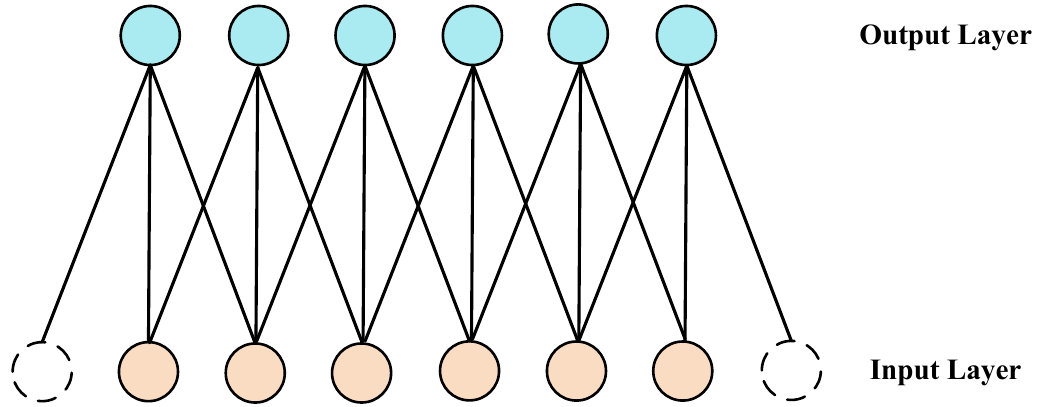}
        \caption{Non-causal convolutional layer.}
    \end{subfigure}
    \caption{Illustration of the mapping relationship of causal and non-causal convolutions, taking a 3$\times$1 convolution kernel with stride = dilation = 1 as an example.}
    \label{fig:two_subfigures}
\end{figure}

\subsection{Causal Convolution}
In scenarios requiring extremely high real-time performance, the latency introduced by models must be strictly constrained. However, for ordinary convolutions, the receptive field of the convolution kernel can access future information, which introduces model latency. For example, for a standard convolution with kernel size $k$ and dilation $d$ , the number of future input frames required is

\begin{equation}
    \zeta (k, d) = \left\lfloor  \frac{(k-1)d}{2}\right\rfloor ,
\end{equation}
where $\left\lfloor  \cdot \right\rfloor$denotes flooring. 
Most GAN-based neural vocoders, structured as stacks of non-causal convolutional layers, exhibit increasing model latency with deeper architectures, which poses certain challenges for real-time applications.

Causal convolutions layers, as illustrated in Figure \ref{fig:two_subfigures}, avoid using future information by asymmetrically padding the input sequence $\bm{x}_T = [x_1, x_2, ..., x_T]^\top$. 
Due to asymmetric padding, the last element of the convolution kernel at the current time step $t$ corresponds to the $t$-th element of the input sequence, which ensures that the output value $\bm{y}_T= [y_1, y_2, ..., y_T]^\top$ at time step $t$ depends only on the input $\bm{x}_{\le t}$, forming a fully causal operation.

\subsection{Model Structure}

As illustrated in Figure \ref{fig:1}, 
for both amplitude and phase prediction in the DLL-APNet vocoder, causal convolutions are first employed for input mel spectrogram, followed by $K$ causal ConvNeXt v2 \cite{woo2023ConvNeXt} blocks for deep feature extraction. 
The original ConvNeXt v2 block consists of a depth-wise convolution with a large kernel, two point-wise convolutions, along with normalization layers and activation functions inserted between them.
Since the point-wise convolutional layers are linear and do not introduce latency, we only convert the original depth-wise convolution into a causal convolution with the same kernel size to minimize model latency, and named this module as causal ConvNeXt v2 block. 
At the output of the amplitude prediction branch, the output of the last causal ConvNeXt v2 block undergoes post-processing via causal convolutions to yield the predicted amplitude spectrum. 
In contrast to the amplitude prediction branch, the phase prediction branch employs a causal parallel estimation architecture (PEA) \cite{ai2024low} to enable explicit phase spectrum prediction. 
In causal PEA, two parallel causal convolutional layers are first used to estimate the pseudo-real and pseudo-imaginary components, and then they are activated using the two-argument arctangent function to estimate the wrapped phase directly. 


\begin{table*}[h!]
	\centering
	\caption{Objective evaluation results of the proposed DLL-APNet vocoder and baselines on the VCTK test set.
}\label{tab1}
	\adjustbox{width=0.95\textwidth}{
		\renewcommand{\arraystretch}{0.97}
		\begin{tabular}{l c c c c c c c c}
			\hline
			\hline
    \multirow{2}{*}&{Causality} &{ SNR}&{LAS-RMSE} & { MCD}& { F0-RMSE} &V/UV error&UTMOS&GFLOPS\\ & &  (dB){ $\uparrow$ }&(dB)$\downarrow$&(dB)$\downarrow$&{ {(cent)$\downarrow$}}&(\%)$\downarrow$&$\uparrow$&$\downarrow$\\
   
			\hline
			{Natural}& -&-&-&-&-&-&4.04&-\\
			\hline
			{BigVGAN} &\ding{55}&6.42&3.63&0.90&21.04&3.22&3.97&230.51\\

			HiFi-GAN& \ding{55}&4.15&4.51&1.58&31.61&3.97&3.93&25.65\\
			{iSTFTNet} &\ding{55}&4.15&5.24&1.87&32.87&4.13&3.93&19.22 \\
   		{APNet2} & \ding{55}&6.56&4.23&0.99&17.38&2.88&4.00&6.30 \\
   		{Vocos} & \ding{55}&6.05&3.70&0.80&25.17&3.47&3.91&2.70\\
			\hline
			causal HiFi-GAN &\ding{51}&3.00&5.24&2.32&58.06&5.96&3.88&25.66\\
			causal {iSTFTNet} &\ding{51}&2.22&5.84&2.27&54.43&6.37&3.75&19.23 \\
   		 causal {APNet2}  & \ding{51}&3.63&4.23&1.63&26.06&4.10&3.90&6.30 \\
   		causal {Vocos}& \ding{51}&5.32&4.33&0.89&32.87&4.32&3.87&2.70\\
   \hline
		 {DLL-APNet} & \ding{51}&6.07&4.29&1.04&20.53&3.16&3.98&6.30\\

			\hline
			\hline
	\end{tabular}}
\end{table*}

\subsection{Knowledge-Distillation-based Training Criteria}
\label{ssec:pp}
Knowledge distillation strategy is commonly employed to leverage the superior learning capabilities of a teacher model to guide the generation of intermediate features in a weaker student model, thereby enhancing the latter's performance. 
Specifically, we first trained a non-causal APNet2 vocoder, which serves as the teacher model. 
Given that causal convolutions can only access current and past information, we leverage features extracted by non-causal modules in APNet2, which are capable of utilizing future context, to guide the causal modules in DLL-APNet student model and enable it to implicitly learn temporal dependencies within causal constraints.
As illustrated in Figure \ref{fig:1}, knowledge distillation is performed after both the input convolution and each ConvNeXt v2 block. 
The L1 distance between the intermediate features of the teacher and student models, serves as the knowledge distillation loss. 
Specifically, we define the output of the input convolutional layer and the $k$-th ConvNeXt v2 block ($k=1,\dots,K$) of teacher model (i.e., APNet2) as $\hat{\bm{O}}$ and $\hat{\bm{O}}_k^{CNX}$. 
The outputs of the student model (i.e., DLL-APNet) at corresponding positions are respectively denoted as $\widetilde{\bm{O}}$ and $\widetilde{\bm{O}}_k^{CNX}$, then the knowledge distillation loss can be defined as:
\begin{align}
    \mathcal{L}_{KD} &= \mathbb{E}_{(\hat{\bm{O}},\widetilde{\bm{O}})}\left\| \hat{\bm{O}} - \widetilde{\bm{O}} \right\|_1 + \nonumber \\
    &\quad \sum_{k=1}^{K} \mathbb{E}_{(\hat{\bm{O}}_k^{CNX},\widetilde{\bm{O}}_k^{CNX})} \left\| \hat{\bm{O}}_k^{CNX} - \widetilde{\bm{O}}_k^{CNX} \right\|_1.
\end{align}

We employed multi-resolution discriminator (MRD) \cite{jang2021univnet} and multi-period discriminator (MPD) \cite{kong2020hifi} to supervise the generated waveforms, ensuring the quality of the synthesized waveforms across various scales and frequency bands. 
For the traning loss, we also introduce the amplitude loss $\mathcal{L}_{A}$, phase loss $\mathcal{L}_P$, reconstructed STFT loss $\mathcal{L}_{S}$, and final waveform loss $\mathcal{L}_{W}$ used in APNet2 \cite{du2023apnet2}, and combine them for adversarially  training DLL-APNet, i.e.,

\begin{equation}
\label{eq4}
    \mathcal{L}=\lambda_A\mathcal{L}_{A}+\lambda_P\mathcal{L}_{P}+\lambda_{S}\mathcal{L}_{S}+\lambda_{W}\mathcal{L}_{W}+\lambda_{KD}\mathcal{L}_{KD},
\end{equation}
where $\lambda_P, \lambda_A, \lambda_S$, $\lambda_{W}$, and $\lambda_{KD}$ are hyperparameters.

\section{Experiments Setups}
\label{sec:es}
\subsection{Dataset}
For our experimental setup, the VCTK-0.92 dataset \cite{veaux2016superseded} was utilized, with all speech utterances downsampled to a 16 kHz sampling rate. 
This corpus comprises recordings from 108 English-speaking individuals, totaling approximately 44 hours of speech material.
Regarding data partitioning, we first extracted samples from 100 speakers. Of these, 90\% were randomly allocated to the training subset, while the remaining 10\% constituted the validation set. For the test set, 2,937 utterances were specifically chosen from the remaining 8 unseen speakers, ensuring a disjoint evaluation corpus.
\subsection{Implementation}
For our proposed DLL-APNet vocoder\footnote{Examples of generated speech can be found at our demo page {https://redmist328.github.io/DLL-APNet/}.}, the amplitude and phase spectra were computed using the STFT with a frame length, frame shift, and FFT size of 320, 80, and 1024, respectively. 
The mel spectrogram was extracted with the same configuration, with a dimensionality of 80. 
We set the hyperparameters as $K=8$, $\lambda_P=100$, $\lambda_A=45$, and $\lambda_S=1$ as the configuration in APNet2 \cite{du2023apnet2}. 
$\lambda_{KD}$ was set to 5 in the main experiment and we will discuss its selection in Section \ref{sssec:kd}. 
The model was trained using the AdamW optimizer for up to 0.5 million steps. 

\subsection{Baselines}
We compared DLL-APNet with BigVGAN \cite{lee2022bigvgan}, HiFi-GAN \cite{kong2020hifi}, iSTFTNet \cite{kaneko2022istftnet}, APNet2 \cite{du2023apnet2}, and 
Vocos \cite{siuzdakvocos}. We reproduced the experimental results using the methods described in their original paper under our experimental implementation.
For fair comparison, We also reproduced their causal versions by replace their origin non-causal convolutional layers with causal convolutional layers.


\subsection{Evaluation Metrics}
In the present research, we utilized five objective metrics for evaluating the quality of synthesized speech. These metrics consist of the signal-to-noise ratio (SNR), root mean square error (RMSE) of log amplitude
spectra (LAS-RMSE), mel-cepstrum distortion (MCD), root mean square error of fundamental frequency (F0-RMSE), and voiced/unvoiced (V/UV) error. And we introduced the neural evaluation metric UTMOS \cite{saeki2022utmos} as an objective auditory perception metric. Moreover, the computational complexity of each model was gauged by the floating-point operations (FLOPs) needed for generating 1-second speech.

\section{Results and Analysis}
\label{sec:ra}

\subsection{Main Experimental Results}
Table \ref{tab1} presents the objective experimental results of the proposed DLL-APNet and baseline vocoders evaluated on the test set. 
Comparative analysis between mainstream vocoders and their causal variants reveals that replacing non-causal convolutions with causal counterparts leads to varying degrees of degradation across all speech quality metrics, indicating an inevitable trade-off between low latency and synthesis fidelity. 
Particularly in F0-related metrics (i.e., F0-RMSE and V/UV error), causal models exhibit substantial performance drops compared to their non-causal counterparts, indicating that relying solely on past and current information for prediction may lead to insufficient information capture, thereby affecting pronunciation accuracy.
However, FLOPs measurements remain nearly unchanged after causal transformation of the models, suggesting that this process imposes minimal impact on computational efficiency. 
Notably, the causal variant of BigVGAN failed to converge during training, rendering it incapable of generating intelligible speech, thus we didn't present its results in Table \ref{tab1}. 
This highlights that causal adaptation is not universally applicable to all vocoder architectures.

A comparison between DLL-APNet and causal vocoders reveals that among all causal vocoders, the proposed DLL-APNet significantly outperformed others in most speech quality metrics, validating the effectiveness of our proposed method. 
In Table \ref{tab1}, the causal APNet2 represents the result of training DLL-APNet without using the knowledge distillation strategy (i.e., $\mathcal L_{KD}=0$).
It can be seen that the SNR, F0-RMSE, V/UV error, and UTMOS metrics of the causal APNet2 all showed significant decreases compared to original APNet2. 
After introducing the knowledge distillation training strategy, these metrics all improve to a certain extent and approach the level of the original APNet2, indicating the effectiveness of the knowledge distillation strategy. 
Although DLL-APNet ranks second only to causal Vocos in terms of FLOPS, it surpasses causal Vocos by 0.09 in UTMOS while maintaining significantly lower FLOPS than other vocoders. 
These results provide theoretical justification for deploying DLL-APNet in real-time and resource-constrained scenarios.

When comparing the proposed DLL-APNet with non-causal neural vocoders, its objective speech quality metrics are comparable to those of other vocoders. 
Additionally, while its UTMOS score is second only to APNet2 and higher than those of other vocoders, it maintains relatively low computational complexity. 
This demonstrates that the proposed DLL-APNet delivers speech quality on par with mainstream vocoders, while also providing low latency and computational efficiency.
\begin{figure}
    \centering
    \includegraphics[width=0.78\linewidth]{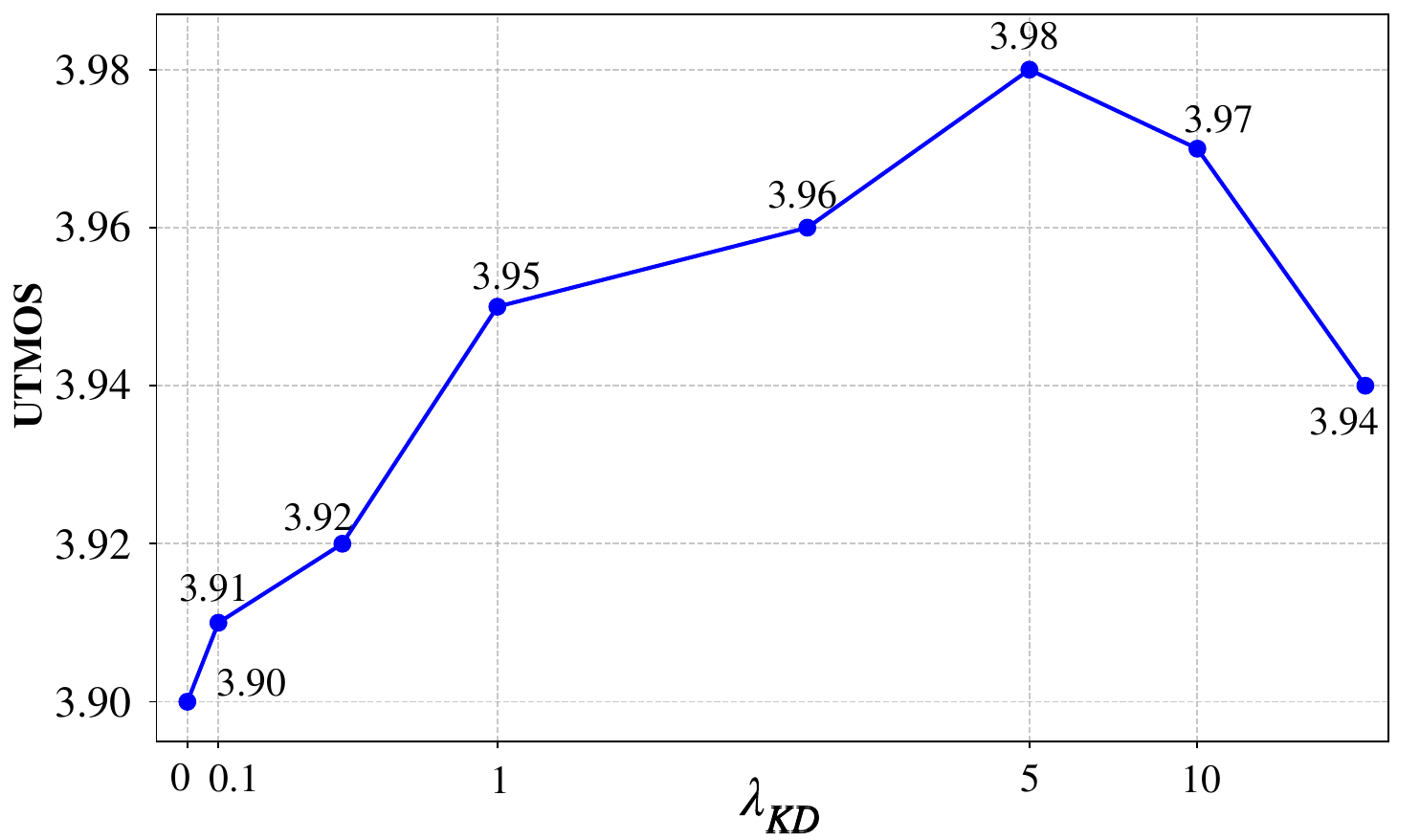}
    \caption{The UTMOS curve of DLL-APNet as the distillation weight $\lambda_{KD}$ varies.}
    \label{fig:3}
\end{figure}
\subsection{Analysis and Discussion}
\label{ssec:eaq}
\subsubsection{Discussion on Distillation Weight Selection}
\label{sssec:kd}
To investigate the impact of the distillation weight $\lambda_{KD}$ on model performance, we conducted a series of controlled-variable experiments on the hyperparameter of knowledge distillation. 
Specifically, we set $\lambda_{KD}$ to 0 (meaning no knowledge distillation strategy is used), 0.1, 0.5, 1, 2, 5, 10, and 20, and explored the UTMOS performance of the DLL-APNet. 
The results are shown in Figure \ref{fig:3}. 
It can be seen that when the hyperparameter value is small, UTMOS increases as the hyperparameter value increases, indicating the effectiveness of the knowledge distillation strategy we used. When the hyperparameter reaches 5, UTMOS reaches the maximum value, and then decreases as the hyperparameter increases. This suggests that $\lambda_{KD}$ should be appropriately set during training. 
Too small values would prevent the knowledge distillation strategy from functioning effectively, while excessively large values could cause the model to overly focus on the distillation loss, neglecting other important objectives such as direct supervision from natural speech, ultimately degrading speech quality.

\subsubsection{Discussion on Distillation Position Selection}

To investigate the impact of distillation positions on model performance of DLL-APNet, we designed analytical experiments by varying the number of distilled layers in the backbone ConvNeXt v2 blocks. 
Specifically, in our setup with 8 ConvNeXt v2 blocks, we trained models by varying the number of blocks involved in knowledge distillation to 0, 2, 4, 6, and 8, respectively. 
The UTMOS results are shown in Table \ref{tab2}. 
It can be observed that speech quality increased with the number of involved blocks, indicating that the effect of knowledge distillation enhanced as the model's participation degree increased. 
These findings also provide useful guidance for the design of knowledge distillation strategies in other speech generation tasks.
\begin{table}[t]
	\centering
	\caption{UTMOS results of DLL-APNet with different distilled numbers of ConvNeXt v2 blocks.
}\label{tab2}
	\adjustbox{width=0.475\textwidth}{
		\renewcommand{\arraystretch}{1.2}
		\begin{tabular}{l c c c c c}
  \hline
  \hline
 Distilled Numbers& 0& 2& 4& 6& 8\\
 \hline
     UTMOS& 3.73& 3.92& 3.95& 3.96& 3.98\\
    
    \hline
    \hline
	\end{tabular}}
\end{table}
\section{Conclusion}
\label{sec:c}
This paper presents a novel causal low-latency neural vocoder, DLL-APNet, which explicitly predicts amplitude and phase spectra from input mel spectrogram using causal-convolution-based predictors and reconstructs speech waveform via iSTFT. 
To enhance speech quality under causality constraints, we employ a pre-trained non-causal APNet2 vocoder as the teacher model to guide the intermediate feature generation of the proposed DLL-APNet vocoder. 
Experimental results demonstrate that DLL-APNet synthesizes speech with higher quality than other causal vocoders and comparable to mainstream non-causal vocoders while requiring fewer computations. 
Future work will focus on reducing model size and computational overhead to better suit practical vocoder applications.

\printbibliography

\end{document}